\documentclass[nofootinbib,amsmath,twocolumn,notitlepage,preprintnumbers,floatfix]{revtex4-1}
\usepackage{multirow}
\usepackage{amssymb, esvect, amsmath, graphicx, latexsym, amsthm, slashed, eso-pic}
\usepackage{xcolor}
\usepackage{amsmath,amsthm,amsfonts,amscd,amssymb} 
\usepackage{eucal}
\usepackage{braket}
\usepackage{slashed}
\usepackage{hyperref}
\usepackage{graphicx}
\usepackage{bm}
\usepackage{siunitx}
\usepackage{xcolor}
\usepackage{afterpage}
\usepackage{changepage}
\usepackage{soul}
\usepackage{float}
\hypersetup{
    colorlinks=true,
    linkcolor=blue,
    citecolor=magenta,      
    urlcolor=cyan}

\usepackage{lipsum}

\usepackage{hyperref}

\usepackage[english]{babel}

\begin{document}

\title{Combining quantum noise reduction resources: a practical approach}

\author{Sohitri Ghosh$^{a,b}$}
\thanks{sohitri@umd.edu}

\author{Matthew A. Feldman$^{c,d}$}
\thanks{feldmanma@ornl.gov}
\author{Seongjin Hong$^{e}$}

\author{Claire E. Marvinney$^{c,d}$}
\author{Alberto M. Marino$^{c,d}$}
\author{Raphael C. Pooser$^{c,d}$}

\author{Jacob M. Taylor$^{a,b}$}
\thanks{jmtaylor@umd.edu; current address Riverlane Research, Inc., One Broadway, Cambridge MA 02139}

\medskip

\affiliation{$^a$Joint Quantum Institute/Joint Center for Quantum Information and Computer Science, University of Maryland, College Park/National Institute of Standards and Technology, Gaithersburg, MD, USA}
\affiliation{$^b$Department of Physics, University of Maryland, College Park, MD, USA}
\affiliation{$^c$Quantum Information Science Section, Oak Ridge National Laboratory, Oak Ridge, TN 37831, USA }\thanks{This manuscript has been authored, in part, by UT-Battelle LLC, under contract DE-AC05-00OR22725 with the U.S. Department of Energy (DOE). The publisher acknowledges the U.S. government license to provide public access under the DOE Public Access Plan (http://energy.gov/downloads/doe-publicaccess-plan).}
\affiliation{$^d$Quantum Science Center,  Oak Ridge National Laboratory, Oak Ridge, TN 37831, USA }
\affiliation{$^e$ Physics Division,  Oak Ridge National Laboratory, Oak Ridge, TN 37831, USA }


\begin{abstract}
Optomechanical sensors are capable of transducing external perturbations to resolvable optical signals. A particular regime of interest is that of high-bandwidth force detection, where an impulse is delivered to the system over a short period of time. Exceedingly sensitive impulse detection has been proposed to observe very weak signals like those due to long range interactions with dark matter that require much higher sensitivities than current sensors can provide. Quantum resources to go beyond the traditional standard quantum limit of these sensors include squeezing of the light used to transduce the signal, backaction evasion by measuring the optimal quadrature, and quantum non-demolition (QND) measurements that reduce  backaction directly. These methods have been developed in the context of gravitational wave detection for target frequencies in the audio band range. Here, we provide the theoretical limits to quantum noise reduction for higher and broader frequency targets, such as those from dark matter signals, while combining quantum enhanced readout techniques based on squeezed light and QND measurements with optomechanical sensors. We demonstrate that backaction evasion through QND techniques dramatically reduces the technical challenges presented when using squeezed light for broadband force detection, paving the way for combining multiple quantum noise reduction techniques for enhanced sensitivity in the context of impulse metrology.

\end{abstract}

\maketitle












\section{Introduction}

In the evolving landscape of physics, the detection of dark matter and other ultra-weak fields and particles represents a paramount challenge, driving the need for innovative and more sensitive detection methodologies. One key approach is measuring the effects of these fields and/or particles using mechanical sensors~\cite{carney2021mechanical}. At this time, four decades of progress in low frequency measurement of mechanical systems for LIGO and other scientific directions has laid the foundation for going beyond the standard quantum limit (SQL) in such sensing domains~\cite{caves1981quantum, jaekel1990quantum, abramovici1992ligo, abbott2004detector, vahlbruch2006coherent, schnabel2010quantum, ligo2011gravitational, aasi2013enhanced, aasi2015advanced, mcculler2020frequency, ganapathy2023broadband}. However, the challenge of leveraging these advances in the impulse or high frequency detection regime, which we focus on here, starts to diverge from a prior focus on ever-better position metrology. 
Our study leverages the pioneering approach of Braginski~\cite{braginsky1980quantum, braginsky1990gravitational} and others~\cite{kimble2001conversion, purdue2002practical} to combine noise reduction techniques like squeezing and back action evasion in a broadband force sensing setting. 
Crucially, the multi-MHz bandwidth necessary for our application space may enable the realization of these ideas due to lower losses and the relative ease of generating squeezing in this range.

Optomechanical sensors, renowned for their exceptional sensitivity for force measurements~\cite{schreppler2014optically, melcher2014self, xu2014squeezing, mason2019continuous}, lie at the heart of our methodology. Traditionally operating near the SQL in position sensitive measurements, these sensors transform mechanical forces into detectable optical signals, and their precision and sensitivity can be improved through the use of quantum noise reduction techniques ~\cite{caves1980measurement, vyatchanin1995quantum, braginsky2003noise, schreppler2014optically, buchmann2016complex, tsang2010coherent, melcher2014self, xu2014squeezing, wimmer2014coherent, PhysRevA.90.043848, mason2019continuous, mehmet2010demonstration}. Our research is set against the backdrop of the burgeoning field of broadband impulse metrology, which heralds a new chapter in detection of particle physics targets including dark matter detection through the measurement of rapid, minute impulses across a wide frequency spectrum~\cite{carney2020proposal, carney2021mechanical, ghosh2020backaction, 
monteiro2020search, carney2023searches}. The Windchime collaboration and its endeavors to detect heavy dark matter through long-range interactions sets a prime example for the potential of these techniques~\cite{attanasio2022snowmass}. Moreover, the implications of impulse metrology extend beyond astrophysics, promising innovations in low-energy single-photon detection and quantum noise-limited pressure calibrations~\cite{fremerey1982spinning, looney1993measurement, arpornthip2012vacuum, makhalov2016primary, eckel2018challenges}.

Central to our investigation is the strategic utilization of squeezed light and quantum non-demolition (QND) measurements leading to backaction evasion (BAE) to go beyond the SQL in a broadband, high-frequency, regime. Squeezed light has been used to reduce measurement added noise below the SQL by enabling the reduction of the uncertainty in the desired variable at the expense of increasing it in its conjugate variable~\cite{pooser2020truncated, pooser2015ultrasensitive, tse2019quantum}. On the other hand, a QND measurement is accomplished when an observable remains unaffected over time due to the quantum uncertainty produced in the corresponding non-commutative conjugate variable, an example being the momentum measurement of the free particle system~\cite{ braginsky1980quantum, caves1980measurement, grangier1998quantum, clerk2008back}. These techniques, primarily advanced by the gravitational wave detection community, have found application in displacement sensing within the audio frequency range~\cite{kimble2001conversion, purdue2002practical}. Our focus, however, extends to high-frequency signals, where it is necessary to address the challenge of backaction noise in scenarios that require substantial laser power to reach the optimum noise floor. The synergy between squeezing and QND momentum measurement presents a significant quantum advantage in reducing quantum noise across a broad frequency band, thus enhancing the sensitivity to weak impulses while also simplifying some of the technical complexities associated with implementing squeezed light, such as not requiring frequency dependent optimization of the  squeezing angle, for the purpose of impulse metrology~\cite{pooser2020truncated, pooser2015ultrasensitive, tse2019quantum, lee2020squeezed, motazedifard2016force}.


This article is structured as follows. We first introduce a toy model to illustrate the advantages of position measurements using single- and two-mode squeezed light. The expected losses in the single-mode case and asymmetry in displacement amplitude of the optical fields in the two-mode case are characterized for the respective cases. Our scheme is shown to be robust against small amounts of loss and displacement amplitude asymmetries. Second, we examine a specific interferometric optomechanical system and find that, in principle,  BAE and squeezing can be combined over a broad bandwidth when reading out the momentum of a resonator well above its natural frequency, thus harnessing the benefits of the QND readout. Our results show that squeezed momentum measurements facilitate broadband impulse metrology with quantum-enhanced sensitivity in the parameter regimes required for dark matter detection. 


\section{Benefits of Squeezing}
\subsection{Single-Mode Squeezing Toy Model}

 \begin{figure*}[ht!]
    \centering
    \includegraphics[scale=0.75]{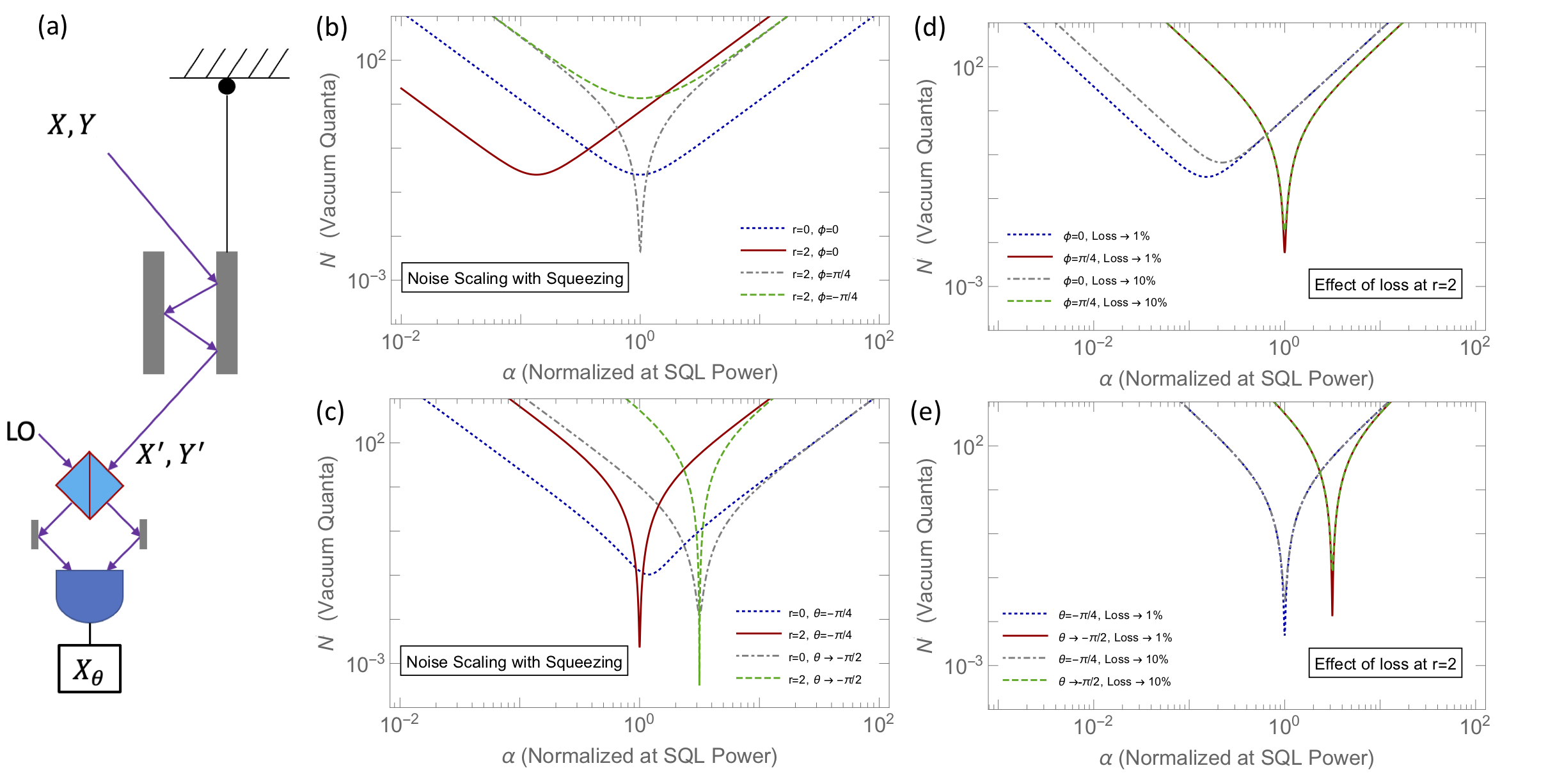}

    \caption{Position measurements using single-mode squeezed light. (a) A position measurement comprising a single-sided optomechanical system probed with single-mode squeezed light characterized by quadratures $X$ and $Y$. In the proposed scheme the measured quadrature $X_\theta$  is read out using homodyne detection by interfering the output with a local oscillator (LO) on a beamsplitter (blue cube) and read out on a balance detector (blue device). (b) Phase quadrature measurement ($\theta = 0$) at squeezing strengths $r=\{0,2\}$ and squeezing angles $\phi =\{0,\pm \pi/4\}$ with a parameter  $\beta = 1$. The displacement amplitude $\alpha$ is normalized to the power  at the SQL, which  is given by the solution to Eq.~(\ref{Noise_single_mode}) when we optimize with respect to $\alpha$ for $\theta, \phi = 0$.  The optimal noise floor for measuring the phase quadrature with $r>0$ occurs at $\phi=\pi/4$ while at $\phi=0$ squeezing provides a reduced power to attain the same noise floor. (c) Noise for measurement quadratures $\theta = \{-  \pi/4,  - \pi/2\}$ with $\phi=0$ and $\beta =1$.  The optimal noise floor occurs at $\theta=-\pi /2$, which corresponds to an amplitude quadrature measurement. (d) Effect of loss on phase quadrature measurements for $\phi =\{0,\pi/4\}$ . (e) Effect of loss on quadrature measurements for $\theta = \{-  \pi/4,  - \pi/2\}$.
     }
    \label{fig:singleModePos}
\end{figure*}

Here, we present a toy model that includes backaction in the measurement system, where light interacts with the position of the system twice, and the interactions are separated by the free evolution of the system in between, as seen in the schematic diagram of Fig.~\ref{fig:singleModePos}(a). An extension of this simple toy model would include an optical cavity, where the light would interact with the test mass many more times, with the proposed experimental implementation of this model with an optical cavity shown in Fig.~\ref{fig:experimental}(a).
 The interaction Hamiltonian for our toy model is given by 
\begin{equation}
    \begin{split}
        H_{\rm int} &= \hbar G  x X \text{,}
    \end{split}
\end{equation}
where $x$ is the position of the system, $X$ is the amplitude quadrature of the light, and $G$ is the optomechanical coupling strength enhanced by the displacement amplitude of the optical field $\alpha$. A full description of the dependence of this interaction on the displacement amplitude is provided in Appendix~\ref{appendix}. In our model, the radiation pressure of the light alters the momentum of the mechanical system. During the free evolution of the system, this shift in momentum causes a change in the position of the optomechanical system. Incorporating the free evolution in between the two interactions allows us to model the effect of the backaction in the system. In the subsequent interaction, the change in position is transduced to the phase quadrature of the light, $Y$. 

The above dynamics have the following unitary evolution, where we approximate the more common continuous process by the Suzuki-Trotter approximation: %
\begin{equation}
    \begin{split}
        U_{\rm tot} = U_{\rm int}U_{\rm free}U_{\rm int}
        &= e^{-i \zeta x_0 X} e^{-i \beta \frac{p_0^{ 2}}{2}} e^{-i \zeta x_0 X} \text{,}
    \end{split}
    \label{unitary_operators_total}
\end{equation}
where $x_0$ and $p_0$ are the dimensionless position and momentum operators of the optomechanical system. These dimensionless variables are normalized to a length and momentum scale comparable to the de Broglie wavelength $\lambda = \sqrt{\frac{\hbar^2}{2 m K_B T}}$ where $m$ and $T$ are the mass and temperature of the moving mirror, respectively. For a harmonic oscillator scenario, the length scale is set by $K_B T \rightarrow \hbar \omega$.  Thus, we interpret the parameter $\beta$ as a dimensionless factor accounting for the free evolution  of the system described by $\beta = \hbar t/ (m \lambda^2) $ with free evolution time $t \sim 1/\kappa$,  where $\kappa$ is the cavity linewidth, such as for an optical cavity based experiment as is proposed in Fig.~\ref{fig:experimental}. The parameter $\zeta$ is dependent on the displacement amplitude of the probing light via $\zeta = \alpha \mu$, where $\mu$ is a dimensionless quantity defined in terms of the interaction time and the single photon optomechanical coupling strength in frequency units $g$, such that $\mu \sim g/ \kappa$. As shown in  Appendix~\ref{appendix}, we find that existing optomechanical systems have sufficient single-photon optomechanical coupling strengths and cavity linewidths for our interaction Hamiltonian to be valid~\cite{G_ref2_Nature_Wilson2015, G_ref3_Nature_Rossi2018}. 

By applying the unitaries in Eq.~(\ref{unitary_operators_total}) we find that the momentum, the phase quadrature of the probing light, and the position  operators transform as follows,
\begin{equation}
    \begin{split}
      U_{\rm int}^\dagger p_0 U_{\rm int} &= p_0 - \zeta X\text{,}\\
      U_{\rm int}^\dagger Y U_{\rm int} &= Y - \zeta x_0\text{,}\\
        U_{\rm free}^\dagger x_0 U_{\rm free} &= x_0 + \beta p_0\text{.}
    \end{split}
    \label{unitary_transform}
\end{equation}
The remaining unitary transformations do not change the operators since $X$, $Y$, and $p_0$ commute with $U_{\rm free}$ while $x_0$ and $X$ commute with $U_{\rm int}$. Thus, the transformation of the operators under the full unitary evolution given by
Eq.~(\ref{unitary_operators_total}) can be represented as,
\begin{equation}
    \begin{split}
      Y^\prime=  U_{\rm tot}^\dagger Y U_{\rm tot} &= Y-2 \zeta x_0 -\zeta \beta p_0 + \zeta^2 \beta X,\\
       X^\prime=  U_{\rm tot}^\dagger X U_{\rm tot} &= X,\\
        x^\prime=  U_{\rm tot}^\dagger x U_{\rm tot} &= x_0+\beta p_0- \beta \zeta X,\\
        p^\prime=  U_{\rm tot}^\dagger p U_{\rm tot} &= p_0- 2\zeta X.\\
    \end{split}
    \label{unitary_transform_total}
\end{equation}

Let us consider the arbitrary quadrature $X_\theta$ at the output of the system described by the unitary transformations in Eq.~(\ref{unitary_transform_total}), such that
\begin{equation}
    \begin{split}
        X_\theta &= Y^\prime \cos{\theta} \hspace{2 mm}  + X^\prime \sin{\theta} \hspace{2 mm}\\
        &= \cos{\theta} (Y-2 \zeta x_0 -\zeta \beta p_0 + \zeta^2 \beta X) + \sin{\theta} X\text{,}
    \end{split}
    \label{arbitrary_quadratures}
\end{equation}
where $X^\prime$ and $Y^\prime$ are the output amplitude and phase quadratures of the system (see  Fig.~\ref{fig:singleModePos}(a)), and $\theta$ is the quadrature angle. We see that the quadrature has shot noise terms independent of $\zeta$, signal terms proportional to $\zeta$, and a backaction term proportional to $\zeta^2$. As can be seen, for the right combination of quadratures, the $\sin{\theta} X$ term can cancel the backaction term. Another way to reduce the backaction noise is through quantum non-demolition measurements~\cite{braginsky1980quantum}. In this scenario, we can continuously monitor a QND variable like momentum in a free particle system instead of the position, to reduce or eliminate the backaction term, thereby achieving backaction evasion. In Sec.~\ref{BAE+squeezing} we discuss how to perform a QND measurement to evade backaction noise in detail. 

For the toy model, we restrict our discussion to the continuous monitoring of position to showcase the benefits of squeezing. 
We obtain the estimator variable  $x_E$ for the system position by dividing $X_{\theta}$ by the multiplicative factor $2\zeta \cos \theta $, 
\begin{equation}
    \begin{split}
        x_E &= X_\theta/ (-2 \zeta \cos \theta)\\
        &= - \frac{Y}{2\zeta}+ x_0 +  \frac{\beta p_0}{2} -\frac {\zeta \beta X}{2}  - \frac{\tan{\theta} X}{2 \zeta}.
    \end{split}
    \label{position_estimator}
\end{equation}
From Eq.~(\ref{position_estimator}), we see that a judicious choice of $\theta$ allows us to cancel out the backaction noise from the amplitude quadrature.
 We find that the optimal  $\theta$ is dependent on $\zeta^2$ (which is proportional to power $\alpha^2$). The noise power spectral density (PSD) of the system position
is proportional to the square of the estimator variable, $x_E$. To analyze the measurement-induced noise of our system we evaluate $N = (x_E - (x_0 + \beta p_0/2))^2$, since $N$ is proportional to the variance of the system position, which may be written in terms of the quadratures and their anticommutators  as 
\begin{equation}
    \begin{split}
       N &= \frac{Y^2}{4\zeta^2} + \frac {\zeta^2 \beta^2 X^2}{4}  + \frac{\tan^2{\theta} X^2}{4 \zeta^2}+ \frac{\beta \{X, Y\} }{4} \\
       & +\frac{\tan \theta  \{X, Y\}}{4 \zeta^2} +\frac{\beta \tan \theta X^2}{2}\text{.}
    \end{split}
    \label{Noise_single_mode}
\end{equation}

For squeezed light all the terms in Eq.~(\ref{Noise_single_mode}) are non-zero. 
The vacuum expectation values of the correlators after squeezing can be represented as~\cite{loudon1987squeezed},
\begin{equation}
    \begin{split}
      \braket{X^2} &= \frac{1}{2}\left(e^{2r} \cos^2{\phi} + e^{-2r} \sin^2{\phi}\right)\text{,}\\
      \braket{Y^2} &=\frac{1}{2} \left(e^{-2r} \cos^2{\phi} + e^{2r} \sin^2{\phi}\right)\text{,}\\
      \braket{\{X,Y\}}&= \frac{1}{2}  (e^{-2r} - e^{2r}) \sin{2\phi}\text{,}\\
    \end{split}
\end{equation}
where $r$ and $\phi$ are the squeezing strength and squeezing angle, respectively. 
For a position measurement, where the phase quadrature ($\theta \rightarrow 0$) is measured, the cross-correlator term can be made negative by choosing a suitable squeezing angle and increasing the squeezing strength to reduce the noise floor, $N$ as seen in Fig.~\ref{fig:singleModePos}(b). For a squeezing angle of $\phi=0$, we can access the same noise floor at a lower power by probing the system with squeezed light (red solid curve) rather than a coherent state (blue dashed curve). On the other hand, a squeezing angle of $\phi=\pi/4$ provides a lower noise floor (gray dashed curve) at the original power. This scenario is attained by adding a negative contribution from the  cross-correlator term to the noise. 

Alternatively, we explore combinations of the quadratures that reduces the backaction term to achieve a lower noise floor when using squeezed light, as can bee seen in Fig.~\ref{fig:singleModePos}(c). We may globally minimize the contribution of the measurement-induced noise by optimizing the quadrature angle $\theta$. The optimization at $\phi =0$ gives us
\begin{equation}
\theta \rightarrow - \tan^{-1} {(\zeta^2 \beta)}\text{.}
\end{equation}
Here, we see that the quadrature angle is strongly dependent on the power of the light. Hence, when working deep in the backaction limit where backaction noise is dominating, at high power we are nearly
measuring the amplitude quadrature that is devoid of any information about the system position, and slight fluctuation in power will destroy this benefit of lowering noise by quadrature angle optimization. 

The optical losses in our system are a critical consideration when probing it with squeezed light. Here, we evaluate the effect of the expected optical losses on the noise floor.  For an overcoupled cavity, as the one considered here, where the intrinsic losses through any channel other than the input port are negligible, the cavity losses would approximately be of the same order as the loss at the detection port. Thus, here we investigate only the effects of the loss at the detection port.  We determine the effect of adding vacuum noise at the output port of the beamsplitter used in homodyne detection. This can be represented by the following phase and amplitude measurement output,
\begin{equation}
    \begin{split}
        Y_{\rm out} &=  Y^{\prime} \cos({\eta}) +  Y^{\prime}_{\rm in} \sin({\eta})\text{,}\\
         X_{\rm out} &=  X^{\prime} \cos({\eta}) +  X^{\prime}_{\rm in} \sin({\eta})
    \end{split}
\end{equation}
where $\eta^2$ quantifies the loss in the system.  Here $Y_{in}^\prime$ and $X_{in}^\prime$ are the input vacuum noise quadratures into the beamsplitter used to model losses. We now focus on measuring the combination of the modified output phase and amplitude quadratures
\begin{equation}
    \begin{split}
        X_{\theta,{\rm out}} &=  Y_{\rm out} \cos{\theta} +   X_{\rm out} \sin{\theta}\\
    \end{split}\text{.}
\end{equation}
By applying the same noise analysis as for the lossless case, we demonstrate that the benefits of squeezing persist even at optical losses expected in experiments (see Figs.~\ref{fig:singleModePos}(d) and~\ref{fig:singleModePos}(e)).

\subsection{Two-mode Squeezing Toy Model}

\begin{figure*}[ht!]
    \centering
    \includegraphics[scale=0.73]{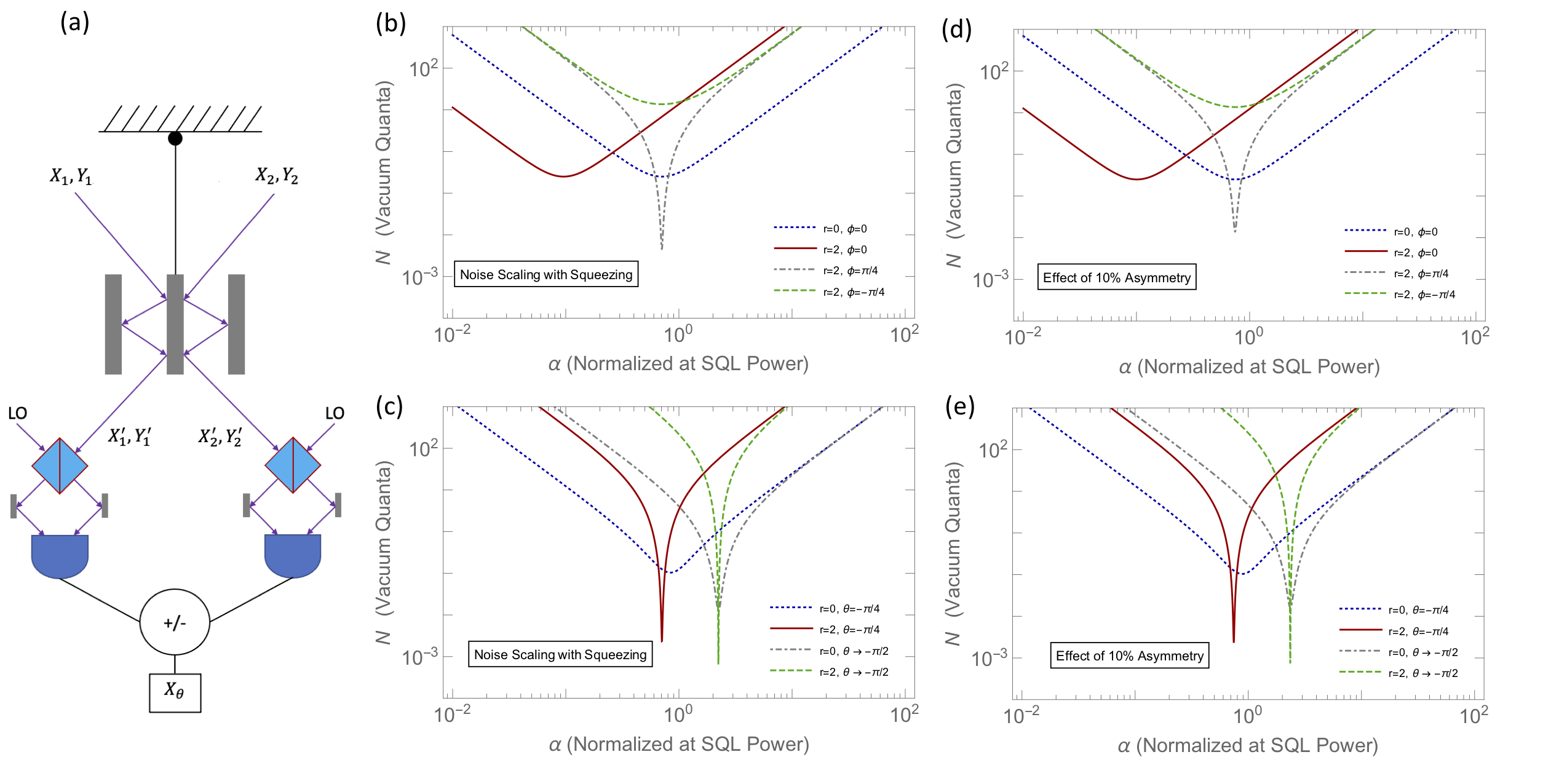}
    \caption{Position measurements using two-mode squeezed light. (a) A position measurement system comprised of a two-sided optomechanical system probed with two-mode squeezed light with quadratures $X_1, X_2$ and $Y_1, Y_2$. In this proposed scheme the measured quadrature $X_\theta$ is read out using homodyne detection by interfering each of the outputs with their respective local oscillators (LO). (b) Phase difference quadrature measurement ($\theta = 0$) at squeezing strengths $r=\{0,2\}$ and squeezing angles $\phi =\{0,\pm \pi/4\}$. The displacement amplitude $\alpha$ is normalized to the power at the SQL for the single-mode (see Fig.~\ref{fig:singleModePos} for comparison). The optimal noise floor for measuring the phase difference quadrature with $r>0$ occurs for $\phi=\pi/4$, while at $\phi=0$ the squeezing provides a reduced power to attain the same noise floor. (c) Noise for measurement quadratures $\theta = \{-  \pi/4,  - \pi/2\}$. The optimal noise floor occurs for $\theta=-\pi /2$, which corresponds to an amplitude difference quadrature measurement. (d) Effect on the phase difference quadrature measurement due to an asymmetry in the power incident on the two-sided cavity with $\alpha_2 = 0.9  \hspace{0.5 mm} \alpha_1$. (e) Effect of power asymmetry on the noise for quadrature measurements $\theta =\{ -  \pi/4,  - \pi/2\}$.
     }
    \label{fig:twoModePos}
\end{figure*}

\begin{figure*}[ht!]
    \centering
    \includegraphics[scale=0.5]{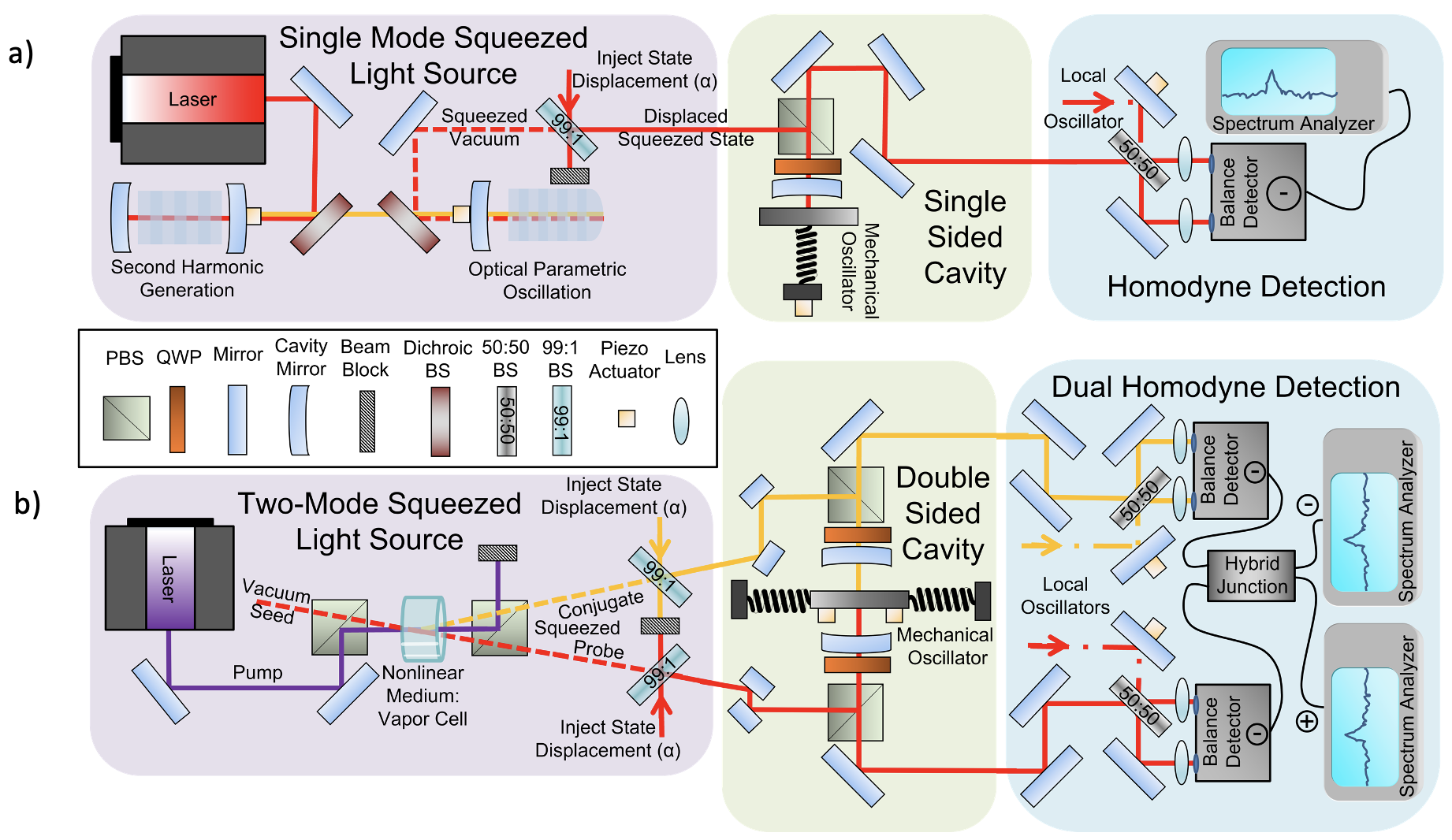}

    \caption{Experimental vision: schematics for position measurement. (a) A single-mode squeezed light source, displaced with an amplitude $\alpha$ after squeezing, is used to interrogate a mirror that can oscillate in an optical cavity. The position of the mirror is encoded in the phase of the probing field and that phase is read out with homodyne detection where a local oscillator is used as a phase reference. (b) A two-mode squeezed light source, with each mode displaced by an amplitude $\alpha$ after squeezing, is used to interrogate an double-sided mirror that can oscillate in a double-sided optical cavity. The position of the mirror is encoded in the phase of the probing field and that phase is read out with a joint measurement of the twin beams using dual homodyne detection, where a pair of local oscillators serve as phase references for the probe and conjugate beams. In the dual homodyne detection scheme, both difference and sum measurements between the joint quadrature modes of the two-mode squeezed light source can be performed. In a practical experiment, an added benefit of probing with the two-mode squeezed light source is the fact that the radiation pressure noise from each beam is equal and opposite to each other. Thus, as in the toy model, the backaction added noise can be tuned by tuning the time delay between the two modes in order to optimize the cancellation of the radiation pressure noise stemming from the correlations in the intensity of the probe and conjugate beams while also improving the signal-to-noise ratio in the measurement.
     }
    \label{fig:experimental}
\end{figure*}

In this section, we consider using two-mode squeezed light to continuously monitor the position of an optomechanical system. We find that for an approximate two-mode interaction, all the benefits expected from the single-mode case can be realized and we show that there is an additional power reduction over that required for the single-mode case. Our system comprises a two-sided optical cavity with a pendulous mirror centered between two fixed mirrors, as seen in the schematic diagram of Fig.~\ref{fig:twoModePos}(a), where the light interacts with the mirror twice in this toy model. The corresponding proposed experimental implementation is shown in Fig.~\ref{fig:experimental}(b), where an actual optical cavity system would be implemented. When the two modes are incident on opposite sides of the pendulous mirror with equal power,  the interaction Hamiltonian takes the form 
\begin{equation}
    \begin{split}
H_{\rm int} = \hbar G x(X_1-X_2)\text{,}
    \end{split}
\end{equation}
where $X_i$ is the amplitude quadrature of light for the $i^{\rm th}$ mode. 

We now evaluate the  toy model shown in Fig.~\ref{fig:twoModePos}(a), where two modes drive our backaction induced
measurement system. 
In a manner analogous to the single-mode case, the dynamics can be characterized by applying the following
chain of unitaries to the system,
\begin{equation}
    \begin{split}
        U_{\rm int}U_{\rm free}U_{\rm int}= e^{-i \zeta x_0 (X_1-X_2)} e^{-i \beta \frac{p_0^2}{2}} e^{-i \zeta x_0 (X_1-X_2)}\text{.}
    \end{split}
\end{equation}
Consequently, the following operators transform to,
\begin{equation}
    \begin{split}
        U_{\rm int}^\dagger p_0 U_{\rm int} &= p_0 - \zeta (X_1-X_2)\text{,}\\
        U_{\rm int}^\dagger Y_1 U_{\rm int} &= Y_1 - \zeta x_0\text{,}\\
         U_{\rm int}^\dagger Y_2 U_{\rm int} &= Y_2 + \zeta x_0\text{,}\\
        U_{\rm free}^\dagger x_0 U_{\rm free} &= x_0 + \beta p_0\text{.}
    \end{split}
    \label{two-mode-operators-evolution}
\end{equation}
Similar to the single-mode case, the remaining transformations do not change the operators since $X_i$, $Y_i$, and $p_0$ commute with $U_{\rm free}$ while $x_0$, and $X_i$ commute with $U_{\rm int}$. 

Under standard homodyne detection schemes, we have access to sums or differences of the quadratures, $X_i$ and $Y_i$ where $i$ denotes the modes. We consider the arbitrary combination of the differences in the $X_i^\prime$ and $Y_i^\prime$ quadratures between the output modes 
\begin{equation}
    \begin{split}
       X_{\theta} &= \cos{\theta} (Y^\prime_1-Y^\prime_2) + \sin{\theta} (X^\prime_1-X^\prime_2)\text{,}\\
        &= (\Delta Y-4\zeta x_0-2\zeta\beta p_0 + 2\zeta^2\beta \Delta X) \cos \theta +\Delta X \sin \theta \text{,}
    \end{split}
    \label{arb_quad_diff}
\end{equation}
where $\Delta Y = Y_1-Y_2$ and $\Delta X = X_1-X_2$ are the phase difference quadrature and amplitude difference quadrature, respectively. Similar to the single-mode case, we find three different contributions to Eq.~(\ref{arb_quad_diff}), shot noise terms that are independent of $\zeta$, signal terms that are proportional to $\zeta$, and backaction term that are proportional to $\zeta^2$.  The estimator variable for measuring the position can once again be obtained by dividing the above quadrature through the multiplicative factor of $x_0$ in Eq.~(\ref{arb_quad_diff}), that is
\begin{equation}
    \begin{split}
        x_E &= X_{\theta}/( - 4 \zeta \cos \theta)\\
        &=  - \frac{1}{4\zeta} \Delta Y +x_0+\frac{\beta p_0}{2}- \frac{\zeta\beta}{2} \Delta X -\frac{\tan \theta}{4\zeta} \Delta X \text{.}
    \end{split}
    \label{position_estimator_two_mode}
\end{equation}

As in the single-mode case, we only consider the measurement-induced noise terms from the light quadratures. Yet again, we can select a $\theta$ in Eq.~(\ref{position_estimator_two_mode}) to negate the $\Delta X$ quadrature from  adding backaction induced noise to the system. We also see that the measurement quadrature angle $\theta$ that minimizes the backaction term is dependent on power ($\alpha^2$). The noise metric $N \propto (x_E - (x_0 + \beta p_0/2))^2$ for the two-mode case is given by
\begin{equation}
N = \frac{1}{16 \zeta^2}(\Delta Y + 2 \zeta^2 \beta \Delta X + \Delta X \tan \theta)^2\text{.}
\end{equation}

We now consider the effect of applying a squeezed two-mode vacuum to our two-sided cavity. For two-mode squeezed light, we see again that we have cross-correlation terms that are non-zero. The relevant vacuum expectation values comprising the amplitude difference and phase difference quadratures for two-mode squeezed light take the form
\begin{equation}
    \begin{split}
      \braket{\Delta X^2} &= e^{2r} \cos^2{\phi} + e^{-2r} \sin^2{\phi}\text{,}\\
      \braket{\Delta Y^2} &= e^{-2r} \cos^2{\phi} + e^{2r} \sin^2{\phi}\text{,}\\
      \braket{\{\Delta X,\Delta Y\}}&=  (e^{-2r} - e^{2r}) \sin{2\phi}\text{,}\\
    \end{split}
\end{equation}
with the individual quadrature correlations 
\begin{equation}
    \begin{split}
      \braket{X_1^2}= \braket{Y_1^2} = \braket{X_2^2}= \braket{Y_2^2}&= \frac{1}{2} \cosh(2r)\text{,}\\
      \braket{X_1Y_1} = \braket{X_2Y_2} &= 0\text{,}\\
       \braket{X_1Y_2} = \braket{X_2Y_1} &=  \frac{1}{2}\sinh(2r) \sin(2\phi)\text{,}\\
       \braket{X_1X_2} = -\braket{Y_1Y_2} &= - \frac{1}{2} \sinh(2r) \cos(2\phi)\text{.}\\
    \end{split}
\end{equation}
From the above correlations we observe that the cross-terms can be made negative by choosing an appropriate squeezing angle and boosting the squeezing strength to lower the noise floor for position measurements, as was the case for single-mode squeezed vacuum. In Figs.~\ref{fig:twoModePos}(b) and~\ref{fig:twoModePos}(c) we show that two-mode squeezed light is an alternative squeezing modality that may be used in position measurements providing the same noise floor with the same dependencies on the squeezing parameter $r$, squeezing angle $\phi$, and quadrature angle $\theta$ as in the single-mode case discussed above. In the two-mode case, the modes have equal amplitude or power. Thus, when we compare Fig. \ref{fig:singleModePos}(b) to Fig.~\ref{fig:twoModePos}(b), in the plot we notice that the optimal power requirement per mode has reduced by a factor of half, which is essentially equivalent to the single-mode case with the  total power distributed over both modes. 

We find that, at a  squeezing angle of $\phi=0$, squeezing reduces the power required to access the SQL level noise floor for our system (see  Fig.~\ref{fig:twoModePos}(b)). Similar  to the single-mode case, we see that for a squeezing angle of $\phi=\pi/4$ and a non-zero squeezing strength 
the noise falls below the SQL near the original displacement amplitude $\alpha$ due to the negative contribution from the cross-correlator term. We globally minimize the noise by optimizing the quadrature angle $\theta$ and find an optimal angle $\theta_{opt} \rightarrow - \tan^{-1} {(2\zeta^2 \beta)}$ that is strongly dependent on power, similar to the single-mode case. This suggests that operating at the optimal angle for high power would correspond to measuring the amplitude difference quadrature, which is devoid of any signal.

Two-mode squeezed light sources may have an asymmetry in the power of each mode. Additionally, there may be mismatches in reflection coefficients for the two-sided cavity. As we show, our toy model is robust to small amounts of asymmetries. If we account for a typical power asymmetry between the signal and the idler, which for the case of a parametric process seeded with either the signal or idler corresponds to the gain and the gain minus one,  we can modify the toy model with different coupling parameters $\zeta_1$ and $\zeta_2$ as follows
\begin{equation}
    \begin{split}
        U_{\rm meas}U_{\rm free}U_{\rm meas} &= e^{-i  x_0 (\zeta_1 X_1- \zeta_2 X_2)}\\
        & e^{-i \beta \frac{p_0^2}{2}} e^{-i  x_0 (\zeta_1 X_1- \zeta_2 X_2)}\text{.}
    \end{split}
\end{equation}
 To understand the impact of the asymmetry, we consider a specific example with a squeezing strength $r = 2$, which corresponds to the case considered in Figs.~\ref{fig:twoModePos}(d) and~\ref{fig:twoModePos}(e). We numerically limit  the expected asymmetry of the modes with displacement amplitudes $\alpha_1$ and $\alpha_2$ so that they do not exceed $\alpha_2/\alpha_1 = 0.9$, as shown in Fig.~\ref{fig:twoModePos}(d). This is the expected limit of the asymmetry we may expect from a realistic experimental implementation using two-mode squeezed light. As our example illustrates, the squeezing and quadrature angles may be tuned to partially recover the characteristics of the noise floors of the symmetric case as seen in Figs.~\ref{fig:twoModePos}(d) and~\ref{fig:twoModePos}(e). We expect that as $r$ increases the amount of allowed asymmetry will be reduced, while still maintaining the advantages of the two-mode case. 

 As can be seen, we find that the phase difference and amplitudes difference quadratures of the two-mode squeezed light behave similar to the phase and amplitude quadratures of the single-mode squeezed light. Advantageously, we also find that the optimal power per mode to achieve SQL level noise floor for the two-mode squeezed light is half of the power required for  single-mode squeezed light.

\section{Practical Design: Single-Sided Cavity}
\label{section-single-sided-cavity}
 
With the intuition gained from our toy model, we investigate the benefits of squeezing in an optomechanical cavity, with potential experimental implementations illustrated in Fig.~\ref{fig:experimental}. In this section, we only investigate the single-sided cavity that is presented in Fig.~\ref{fig:experimental}(a), and do not go into the details for the more complicated two-sided cavity that is shown as a possible experiment in Fig.~\ref{fig:experimental}(b). For the single-sided case, the linearized interaction Hamiltonian for the cavity is 
\begin{equation}
 H_{\rm{int}} =  \hbar G x X\text{,}
\end{equation}
where $G$ is the optomechanical coupling strength in frequency per unit length and $x$ is the position of the moving mirror. The Hamiltonian $H_{\rm tot}$ that describes the optomechanical system in the rotating frame of the drive with a detuning $\Delta$ from the cavity resonance is as follows,
\begin{equation}
H_{\text{tot}} = H_{\text{cav}}+H_{\text{mech}}+H_{\rm int}+ H_{\text{bath}}\text{,}
\end{equation}
with
\begin{align}
\begin{split}
H_{\text{cav}}&= -\hbar \Delta a^\dagger a\text{,}\\
H_{\text{mech}}&= \frac{1}{2}m \omega_m^2 x^2+\frac{p^2}{2m}\text{,}
\end{split}
\end{align}
where $H_{\rm cav}$, $H_{\rm mech}$, and $H_{\rm bath}$ are the Hamiltonians for the optical cavity, the mechanical oscillator, and the bath coupled to the system, respectively. The parameters $m$ and $p$ are the mass and momentum of the moving mirror, respectively, in our system and $\omega_m$ is the mechanical resonance frequency of the moving mirror. 

For a drive on resonance with the cavity, $\Delta=0$, the Heisenberg-Langevin equations of motion are
\begin{align}
\begin{split}
\dot{X} &= -\frac{\kappa}{2}X+ \sqrt{\kappa} X_{\rm in}\text{,}\\
\dot{Y} &= -G x -\frac{\kappa}{2}Y+ \sqrt{\kappa} Y_{\rm in}\text{,}\\
\dot{p} &= -\hbar G X - \gamma p +  F_{\rm in}-m\omega_m^2 x\text{,}\\ 
\dot{x} &= \frac{p}{m},
\end{split}
\end{align}
where $\kappa$ is the cavity decay rate, $X_{\rm in}$ and $Y_{\rm in}$ are the quadratures of the input fields to the system, $F_{\rm in}$ is the external force acting on the resonator, $X$ and $Y$ are intra-cavity field quadratures and $\gamma$ is the mechanical damping rate. The bath operators are expressed in terms of the input fields~\cite{RevModPhys.82.1155,gardiner1985input}. The output quadratures are related to the input quadratures by the input-output relations
\begin{align}
\begin{split}
\label{input-output-single}
X_{\text{out}} &= X_{\text{in}} - \sqrt{\kappa} X\text{,}\\
Y_{\text{out}} &= Y_{\text{in}} - \sqrt{\kappa} Y.
\end{split}
\end{align}

We find $X_{\rm out}$ and  $Y_{\rm out}$ by first solving in the frequency domain for variables $X,~Y,~x$, and $p$ in terms of the input fields $X_{\rm in},~Y_{\rm in},$ and $F_{\rm in} $. We then apply the input-output relations to show that 
 \begin{equation}
 \begin{split}
 X_{\rm out} &= e^{i \phi_c} X_{\text{in}}\text{,}\\
Y_{\text{out}} &= e^{i \phi_c} Y_{\text{in}} + G  \chi_c \chi_m \left[F_{\text{in}} -\hbar G \chi_c X_{\text{in}} \right],
\end{split}
\label{Xout_and_Yout_pos}
\end{equation}
where the cavity response function $\chi_c$, the mechanical response function $\chi_m$, and the cavity phase shift $e^{i \phi_c}$ are given by
\begin{equation}
\begin{split}
\chi_c &= \frac{\sqrt{\kappa}}{-i \nu+\kappa/2} \text{,}\\ 
  \chi_m &= \frac{-1}{m(\nu^2-\omega_m^2 +i\gamma \nu)} \text{,}\\
e^{i \phi_c} &= \frac{-i \nu - \kappa/2}{-i \nu + \kappa/2}\text{.}
\end{split}
\end{equation}

We define the measurement quadrature for the output field from the system as
\begin{equation}
    \begin{split}
        X_{\theta} &= Y_{\rm out} \cos{\theta}  + X_{\rm out} \sin{\theta} .
    \end{split}
    \label{arbitrary_quad_pos}
\end{equation} 
By inspection of Eqs.~(\ref{Xout_and_Yout_pos}) and~(\ref{arbitrary_quad_pos}) we obtain the estimator variable for the input force, $F_E$, by dividing the output quadrature by the coefficient of the input force term  $G\chi_c\chi_m \cos\theta$,
\begin{equation}
\begin{split}
F_E &= \frac{ X_\theta }{G   \chi_c \chi_m  \cos \theta}\\
&= \left[\frac{e^{i \phi_c} \tan \theta}{G \chi_c \chi_m}- \hbar G  \chi_c \right] X_{\rm in} + \frac{e^{i \phi_c }Y_{\rm in}}{ G\chi_c \chi_m}  + F_{\rm in}.
\end{split}
\end{equation}
We determine the two point correlation function $\braket{F_E (\nu) F_E (\nu^\prime)}$ for $F_E$ since  it is proportional to the noise PSD $S_{ FF} (\nu)$ according to 
\begin{equation}
    \braket{F_E (\nu) F_E (\nu^\prime)} = S_{FF} (\nu) \delta(\nu +\nu^\prime).
\end{equation}

Understanding how to optimize the frequency-dependent power spectral density allows us to facilitate an intelligent search for either a monochromatic or broadband signal. For a monochromatic signal, a frequency-dependent optimization of the PSD is required to obtain the maximum sensitivity. On the other hand, the optimized signal to noise ratio for a broadband signal corresponds to the noise PSD integrated over a bandwidth of interest. For this case, an optimization over a broad frequency range would be necessary to obtain maximum sensitivity.

To consider the different parts of the noise, we focus on the quantity $\braket{F_E^2}$. In the limit in which the mechanical decay rate $\gamma$ is much smaller than the other frequency scales ($\gamma \ll \omega_m, \kappa$), the $\gamma$-dependent imaginary term of the cross-correlator may be neglected to obtain the following expression
\begin{equation}
\begin{split}
\braket{F_E^2} &=  \left|\frac{e^{i \phi_c} \tan \theta}{G \chi_c \chi_m}- G \hbar \chi_c \right|^2 \braket{X_{\rm in}^2 }+\left| \frac{1}{ G\chi_c \chi_m}\right|^2 \braket{Y_{\rm in}^2} \\
& + \left[ \hbar m (\omega_m^2-\nu^2) +\frac{\tan \theta}{G^2 |\chi_c|^2 |\chi_m|^2} \right] \braket{\{X_{\rm in}, Y_{\rm in}\}} \\
& + \braket{F_{\rm in}^2}.
\end{split}
\label{eqn_27}
\end{equation}
To optimally reduce the noise, we minimize $\braket{F_E^2}$ with respect to the quadrature angle $\theta$ to reduce the backaction noise. Prior to minimizing the noise, 
we fix the squeezing angle $\phi$ to zero. 
This leads to the optimized quadrature angle 
\begin{equation}
\theta_{\rm opt} \rightarrow \tan^{-1} \left[ {\hbar G^2 m |\chi_c|^2 |\chi_m|^2 (\nu^2-\omega_m^2)}\right].
\label{Angle_pos}
\end{equation} 
By inspecting Eq.~(\ref{Angle_pos}) we see that $\theta_{\rm opt}$ is strongly dependent on power ($G^2 \propto \alpha^2$) and frequency. 
In the limit of high power and low frequency, $\theta_{\rm opt}=\pi/2$. This corresponds to measuring the amplitude quadrature $X$, which is devoid of any signal from $F_{\rm in}$. However, in the high frequency limit $\theta_{\rm opt}$ approaches 0, making it possible to measure a signal at an optimally low noise floor.

Our objective is to minimize the terms of $\braket{F_E^2}$ that correspond to measurement-induced noise, $N(\nu)$, thus we only consider $N(\nu)$ for the remainder of the section. While the thermal contribution to the noise may be reduced as discussed in~\cite{carney2020proposal}, it is not the focus of this article. At the optimal quadrature angle and in the low $\gamma$ limit, the measurement-induced noise $N(\nu)$ for an arbitrary squeezing angle $\phi$ simplifies to
\begin{equation}
\begin{split}
N(\nu) = \frac{e^{-2r} \cos^2{\phi} + e^{2r} \sin^2{\phi}}{2 G^2 |\chi_c|^2 |\chi_m|^2}.
\end{split}
\label{quantum_noise_opt1_sect3}
\end{equation}
Only the shot noise term is present in Eq.~(\ref{quantum_noise_opt1_sect3}) since for $\theta \rightarrow \theta_{\rm opt}$ the backaction term of $N(\nu)$ cancels out as $\gamma \rightarrow 0$. Here, the optimal choice for squeezing angle is $\phi =0$. With these parameters the quantum noise is the following, which suggests that we can enhance the power and squeezing strength to bring down the overall noise floor, 
\begin{equation}
\begin{split}
N(\nu) &= \frac{e^{-2r}}{2 G^2 |\chi_c|^2 |\chi_m|^2} \\
&\approx \frac{e^{-2r} m^2 (\kappa^2/4 +\nu^2) (\nu^2-\omega^2)^2}{2 G^2 \kappa}.
\end{split}
\end{equation}

As an alternative to this approach, by restricting ourselves to phase quadrature measurements, we may optimize $N(\nu)$ with respect to power. A power optimization with $\phi=0$ leads to a target frequency-dependent coupling strength $G$, 
\begin{equation}
    G(\nu) \rightarrow \frac{e^{-r}}{\sqrt{\hbar} |\chi_m(\nu)|^{1/2} |\chi_c(\nu)|},
    \label{Optimized G}
\end{equation}
which demonstrates that when squeezing is present and the squeezing angle is zero 
the optimum power required to reach the same noise floor as the SQL is reduced, consistent with our toy model. Moreover, when we measure the phase quadrature while operating at the original optimized coupling strength (using the same power as for the case without squeezing: $r=0$ in Eq.~(\ref{Optimized G})), a contribution to the total noise will be from the sum of the balanced shot noise and backaction noise terms. This sum is equivalent to the SQL. The additional contribution is from the cross-correlator term (the third term in Eq.~(\ref{eqn_27})). Careful selection of the squeezing angle $\phi$, will force the cross-correlator term to negatively contribute to the total noise expression, thus reducing the overall noise. Under these conditions, in the $\gamma \rightarrow 0$ limit, the total measurement-induced noise takes the form
\begin{equation}
\begin{split}
N(\nu) =\hbar m (\nu^2-\omega_m^2) (\cosh 2r+ \sinh 2r \sin 2 \phi).
\end{split}
\label{33_Nquantum}
\end{equation}
 Equation~(\ref{33_Nquantum}) shows that the measurement-induced noise is minimized by squeezing when the squeezing angle is tuned to $\phi = -\pi/4 $.


\section{Combining Backaction Evasion and Squeezing}
\label{BAE+squeezing}
\subsection{Continuous Momentum Measurement}

In this section we consider a backaction evading model achieved by approaching a quantum non-demolition measurement of momentum, where we continuously monitor the momentum rather than the position of an optomechanical system using single-mode squeezed light.
Momentum is effectively a QND variable while working in the free particle limit~\cite{braginsky1980quantum}. 

 \begin{figure}[ht!]
    \centering
    \includegraphics[scale=0.4]{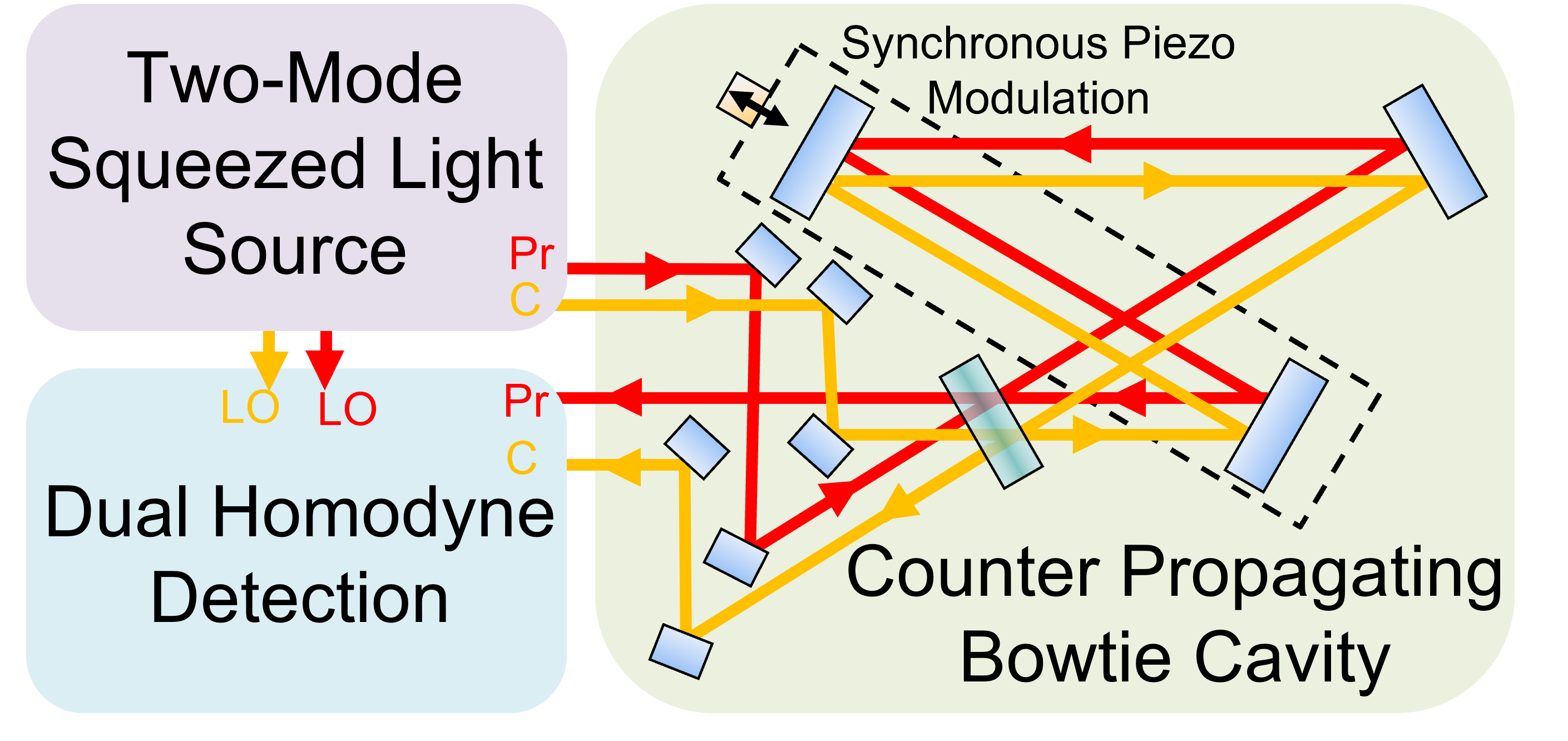}
    \caption{Momentum measurement configuration in a bowtie cavity probed by  counter-propagating beams of a two-mode squeezed state. The modes enter and exit the cavity through a beam-splitter. The momentum-sensitive measurement stems from the synchronous modulation of a pair of the cavity mirrors, which do not change in position with respect to each other, but do have a shared velocity. Thus, the velocity and the momentum of those synchronously driven mirrors can be determined when the pair of  modes (Pr: probe; C: conjugate) that interact with the bowtie cavity are interfered with their respective local oscillators (LOs) in a dual-homodyne detection scheme (as in 
    Fig.~\ref{fig:experimental}(b)), where a difference measurement or sum measurement of the joint quadratures of the two-mode squeezed state is accessible. Additionally, it is possible to overlap the spatial modes of the counter propagating twin beams by cross-polarizing the modes and using polarizing beamsplitters for injecting and collecting light from the cavity.}
    \label{fig:bowtie-cavity}
\end{figure}

In our model, the light quadratures interact directly with the mechanical momentum, which is proportional to the velocity of the movable mirror in the system rather than its position. Thus, the interaction Hamiltonian is dependent on the momentum $p$ of the mechanical system, 
\begin{equation}
  H_{\rm int}= \hbar  G^\prime p X,
  \label{momentum_int}
\end{equation}
where $G^\prime$ is the optomechanical coupling strength in units of frequency over momentum.
While this interaction cannot be directly generated by the single-sided optomechanical cavity discussed in the last section, there are practical designs of optomechanical systems, such as those described in~\cite{ghosh2020backaction}, as well as electromechanical couplings \cite{richman2023general} that showcase the type of interaction described in Eq.~(\ref{momentum_int}). Furthermore, one possible system for a velocity measurement is a bowtie cavity, for which the two-mode squeezed light can be used  to enable an experimental realization for a momentum measurement, as shown schematically in Fig.~\ref{fig:bowtie-cavity}, with the modes of the two-mode squeezed light counter propagating through a bowtie cavity with a pair of synchronously driven mirrors.  The position of the mirrors with respect to each other does not change; however, the mirrors experience a velocity that does not cancel out and thus can be observed. We show that this velocity- and thus momentum-based measurement
combined with squeezed light demonstrates a method for combining shot noise reduction below the SQL with a QND measurement. 

We focus the remainder of this section on the benefits of a momentum based interaction and do not consider its practical implementation any further. For the interaction Hamiltonian given in Eq.~(\ref{momentum_int}), the Heisenberg-Langevin equations of motion take the form
\begin{align}
\begin{split}
\dot{X} &= -\frac{\kappa}{2}X+ \sqrt{\kappa} X_{\rm in}\text{, }\\
\dot{Y} &= -G^\prime  p -\frac{\kappa}{2}Y+ \sqrt{\kappa} Y_{\rm in}\text{, }\\
\dot{p} &=  - \gamma p +  F_{\rm in}-m\omega_m^2 x\text{,}\\ 
\dot{x} &= \frac{p}{m}+\hbar G^\prime X.
\end{split}
\end{align}
The input and output quadratures for the optical fields are related by the input-output relations~\cite{RevModPhys.82.1155,gardiner1985input}
\begin{align}
\begin{split}
\label{input-output-momentum}
X_{\text{out}} &= X_{\text{in}} - \sqrt{\kappa} X\text{, and}\\
Y_{\text{out}} &= Y_{\text{in}} - \sqrt{\kappa} Y.
\end{split}
\end{align}
We solve for the 
output quadratures in terms of the input quadratures 
as was done to obtain Eq.~(\ref{Xout_and_Yout_pos}) in the previous section. We find that they are given by 
 \begin{equation}
 \begin{split}
 X_{\rm out} &= e^{i \phi_c} X_{\text{in}}\text{,}\\
Y_{\text{out}} &= e^{i \phi_c} Y_{\text{in}} - i  G^\prime \chi_c \chi_m  m \nu F_{\text{in}} \\
&-\hbar  m^2 \omega_m^2 G^{\prime 2} \chi_c^2 \chi_m X_{\text{in}}.
\end{split}
\label{in_out_p}
\end{equation}

 As in the previous section, to estimate the force acting on the system, we divide a particular combination of the phase and amplitude quadrature as in Eq.~(\ref{arbitrary_quad_pos}) by the multiplicative factor for $F_{\rm in}$,  resulting in the force estimator variable 
 \begin{equation}
\begin{split}
F_E &= \frac{ X_\theta }{- i G^\prime   \chi_c \chi_m m \nu \cos \theta }\text{,}\\
&= \left(\frac{i e^{i \phi_c} \tan \theta}{G^\prime  m \nu  \chi_c \chi_m}- i G^\prime  \hbar \chi_c m \frac{\omega_m^2}{\nu} \right) X_{\rm in}\\
&+ \frac{i e^{i \phi_c}}{ G^\prime m \nu \chi_c \chi_m} Y_{\rm in} + F_{\rm in}.
\end{split}
\end{equation}
Our interest is in the contributions of the measurement-induced noise to the total noise PSD. To investigate these contributions, we determine the two point correlation function of the estimator force variable, as in the previous section, and find that it takes the form
\begin{equation}
\begin{split}
\braket{F_E^2} &=  \left|\frac{i e^{i \phi_c} \tan \theta}{G^\prime  m \nu  \chi_c \chi_m}- i G^\prime \hbar \chi_c m \frac{\omega_m^2}{\nu} \right|^2 \braket{X_{\rm in}^2 }\\
& +\left|\frac{i e^{i \phi_c}}{ G^\prime m \nu \chi_c \chi_m}\right|^2 \braket{Y_{\rm in}^2} + \braket{F_{\rm in}^2}+\\
&  \left[ \hbar m \frac{\omega_m^2}{\nu^2}(\omega_m^2-\nu^2) +\frac{\tan \theta}{m^2 \nu^2 G^{\prime 2}|\chi_c|^2 |\chi_m|^2} \right]\braket{\{X_{\rm in}, Y_{\rm in}\}}.
\end{split}
\label{two_point_corr_fun_momentum}
\end{equation}
Note that there is no backaction noise which is directly proportional to the coupling strength $G^\prime$, in the limit where the mechanical frequency  $\omega_m \rightarrow 0$, indicating the system is in the free particle limit. 

For $\phi =0$, we find from Eq.~(\ref{two_point_corr_fun_momentum}) that the optimal quadrature angle that minimizes the noise is given by
\begin{equation}
\theta_{opt} \rightarrow \tan^{-1} \left[ {\hbar G^{\prime^2} m^3 \omega_m^2 |\chi_c|^2 |\chi_m|^2 (\nu^2-\omega_m^2)}\right].
\label{Angle_vel}
\end{equation}
In the limit $\omega_m \rightarrow 0$, the noise is minimized by measuring the phase quadrature, which corresponds to $\theta_{\rm opt}=0$. Under these conditions, the backaction term is canceled, leaving only the shot noise term to contribute to the measurement-induced noise. Thus, as $\theta \rightarrow \theta_{\rm opt}$, in the low $\gamma$ and free particle limit, the measurement-induced noise takes the form
\begin{equation}
\begin{split}
N(\nu) = \frac{e^{-2r} \nu^2 (\cos^2 \phi+ e^{4r} \sin^2 \phi )}{2 G^{\prime 2} |\chi_c|^2 }.
\end{split}
\label{quantum_noise_momentum2}
\end{equation}

Since $G^{\prime 2}$ is proportional to power, an increase in power will monotonically lower the noise floor. Squeezing can further minimize the measurement-induced noise for momentum sensing. By inspecting Eq.~(\ref{quantum_noise_momentum2}), we find that in this scenario, the optimal squeezing angle to lower the measurement-induced noise is $\phi = 0$. As in the case of position sensing, if we focus on the alternative approach where we only measure the phase quadrature, we can also optimize the measurement-induced noise with respect to power and obtain a target frequency-dependent coupling strength, which in this case takes the form 
\begin{equation}
    G^\prime(\nu) \rightarrow \frac{e^{-r}}{m \omega_m \sqrt{\hbar} |\chi_m(\nu)|^{1/2} |\chi_c(\nu)|}.
    \label{Optimized Gprime}
\end{equation}
This result shows that  squeezing lowers the necessary power to attain a given noise floor. 

Using the optimized power level when squeezing is not used in the total noise expression given by Eq.~(\ref{two_point_corr_fun_momentum}), we obtain the noise contributions from the SQL equivalent noise and the cross-correlator terms.
Similar to the toy model and the position sensing scheme presented above, we see that by tuning the squeezing angle to $\phi=-\pi/4$, we force the cross-correlator term to become negative, thereby reducing the total noise. In the $\gamma \rightarrow 0$ limit, the measurement-induced noise expression takes the form
\begin{equation}
\begin{split}
N(\nu) =\hbar m \frac{\omega_m^2}{\nu^2}(\nu^2-\omega_m^2) (\cosh 2r+ \sinh 2r \sin 2 \phi).
\end{split}
\end{equation}

\subsection{Comparisons}

Here, we compare the sensitivity of the force measurements that can be achieved in the position sensing regime to that of those achieved in the momentum sensing regime, aimed towards different types of signals.  For position sensing, the coupling strength $G$ scales as the frequency over the cavity length; while for momentum sensing, $G^\prime$ scales as the frequency over momentum. For each technique, during the measurement we coherently integrate the signal over the cavity lifetime. The momentum coupling strength $G^\prime$ is then related to $G$ by the factor $1/ (m \kappa)$ where $m$ is the mass of the mechanical system and $\kappa$ is the cavity decay rate. The reciprocal $1/\kappa$ is also  the natural scale in~\cite{ghosh2020backaction}.

\begin{figure}[t h]
    \centering
    \includegraphics[scale=.97]{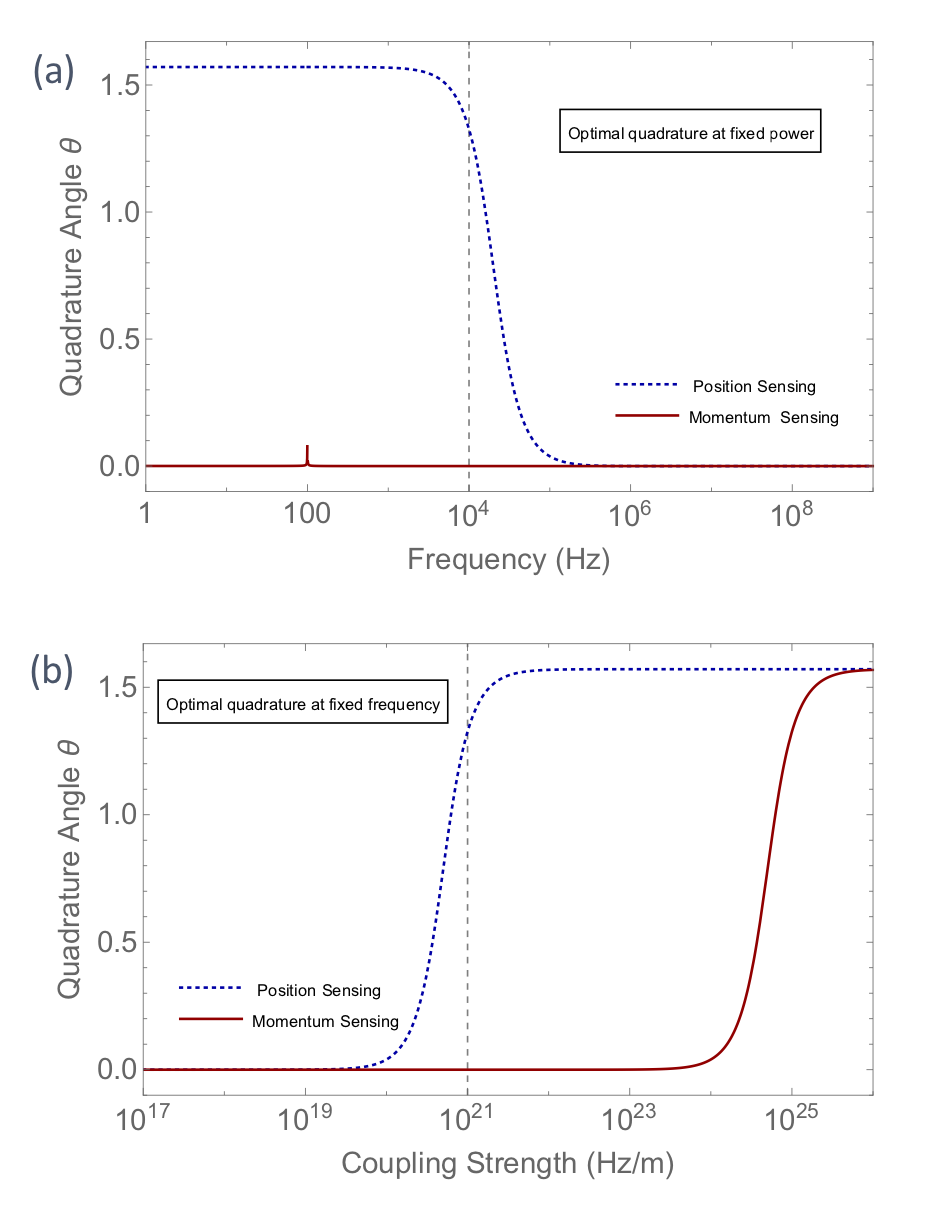}
    \caption{Optimal quadrature angle comparison between position sensing, given by Eq.~(\ref{Angle_pos}), and momentum sensing, given by Eq.~(\ref{Angle_vel}). (a) Optimal quadrature angle plotted as a function of frequency for a fixed optical power. The parameters used are a  mechanical frequency $\omega_m = 100 \text{  Hz}$, a cavity decay rate $\kappa = 1 \text{ MHz}$, a mirror mass $m = 1  \text{ mg}$, a damping rate $\gamma = 0.1 \text{ mHz}$, and  
    a coupling strength $G$ fixed at $10^{21}~{\rm Hz/m}$ with $G^\prime \rightarrow G/ m \kappa$. Position sensing has an optimal measurement angle $|\theta| \rightarrow \pi/2$ (amplitude quadrature) at lower frequencies and $|\theta| \rightarrow 0$ (phase quadrature) at higher frequencies; whereas for momentum sensing $|\theta| \rightarrow 0$  for a broad spectrum. (b) Optimal quadrature angle plotted as a function of power at a fixed frequency of $\nu \rightarrow 10~\text{kHz}$. We see that $|\theta|$  starts to deviate from 0 at much lower powers for position than for momentum sensing.}
    \label{fig:Angle}
\end{figure}

We begin by comparing the optimum quadrature angle that maximizes the signal-to-noise for the position and momentum sensing techniques. For this comparison, the coupling strengths $G$ and $G^\prime$ are chosen to correspond to the same power in both techniques.  In Fig.~\ref{fig:Angle}(a), we see that for a fixed value of $G$ (corresponding to an input power of around 0.1~mW for a 1~cm cavity), the optimal quadrature for position sensing at low frequencies is the amplitude quadrature and at higher frequencies is the phase quadrature. Whereas, under the same conditions, for momentum sensing we find that the optimal quadrature  is the phase quadrature across a broadband frequency spectrum. For a fixed frequency (here $\nu=10~\text{kHz}$) we find that the optimal quadrature for position sensing deviates from the phase quadrature at much lower powers than in does for the momentum sensing, as shown in Fig.~\ref{fig:Angle}(b). 

To compare the performance of the position and momentum sensing protocols, we must first distinguish between the relevant search strategies for specific signals. For dark matter detection, we  distinguish between a monochromatic signal, such as the one from the coherent field generated by ultralight dark matter candidates~\cite{carney2021ultralight}, and a broadband impulse signal, as the one expected for heavy dark matter candidates~\cite{carney2020proposal}. For the monochromatic signal, considering  narrow-band or resonant search strategies is more effective. For a heavy dark matter search including long-range interactions, the target frequency band is 1-10 MHz for the expected short impulse signal; thus, we consider a broadband search.

\begin{figure}[t!]
    \centering
    \includegraphics[scale=0.48]{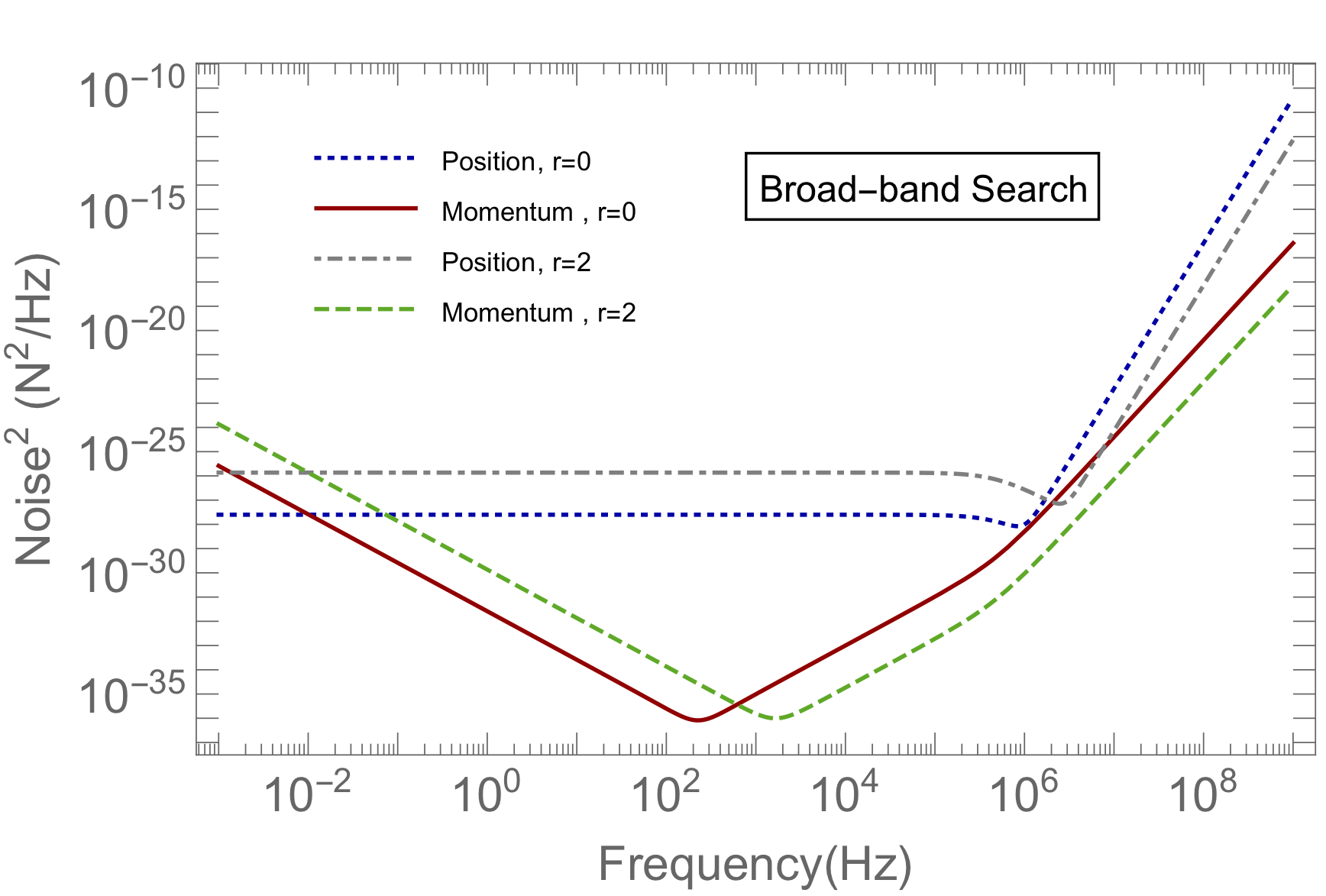}
    \caption{Broadband search strategy for position and momentum sensing. The noise corresponding to the phase quadrature is plotted for both sensing protocols while operating at the optimal power for position sensing with a center frequency of $1 \text{ MHz}$ (using Eq.~(\ref{Optimized G}) for $r=0$, $\nu = 1 \text{ MHz}$). The same power is used for the squeezed scenario. The optomechanical coupling strengths in these techniques are related by $G^\prime \rightarrow G/( m \kappa)$ with a  mechanical frequency $\omega_m = 100 \text{  Hz}$, a cavity decay rate $\kappa = 1 \text{ MHz}$, a mirror mass $m = 1  \text{ mg}$, a damping rate $\gamma = 0.1 \text{ mHz}$, and  a squeezing angle $\phi = 0$. The noise for the momentum sensing is lower than the one for position sensing across a broad frequency range. For a fixed power, squeezing lowers the noise in the shot noise dominated region for both sensing protocols.  }
    \label{fig:broad-band}
\end{figure}

\begin{figure}[t!]
    \centering
    \includegraphics[scale=0.8]{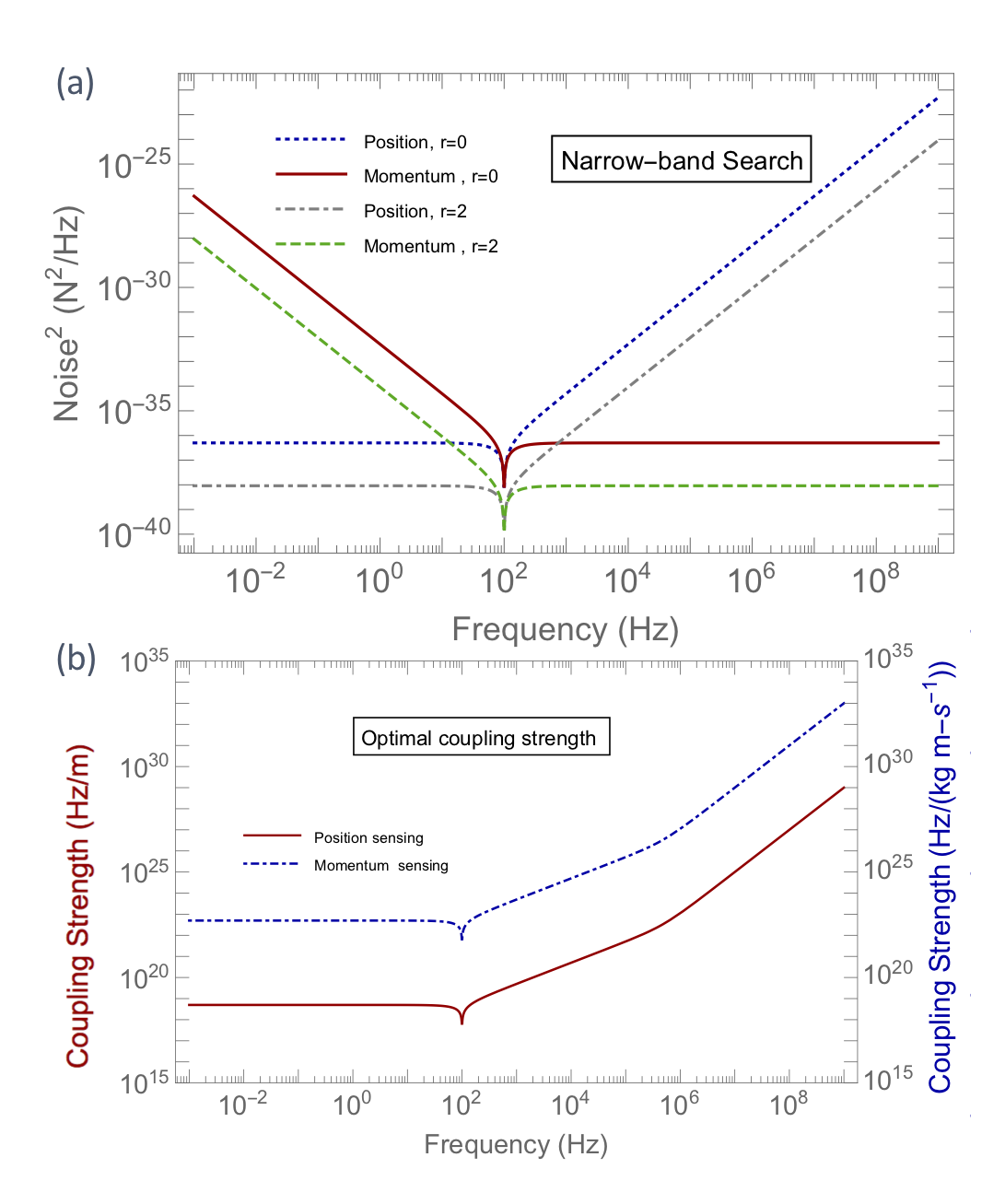}\\
    \caption{(a) Narrow-band search strategy for position and momentum sensing. The noise is plotted for both techniques for an optimal power (Eqs.~(\ref{Optimized G}) and~(\ref{Optimized Gprime}) for $r=0$) and optimal quadrature angle (Eqs.~(\ref{Angle_pos}) and~(\ref{Angle_vel})) for each frequency. To calculate the sensitivities, we consider a mechanical frequency $\omega_m = 100~\text{Hz}$, a cavity decay rate $\kappa = 1 \text{ MHz}$, a sensor mass $m = 1~\text{mg}$, a damping rate $\gamma = 0.1~\text{mHz}$, and a squeezing angle $\phi = 0$. The optimal noise floor is lower for momentum sensing than for position sensing when $\nu > \omega_m$. Squeezing at the same power reduces the noise floor over the full spectrum, except on resonance.  (b) The optimal coupling strength dependence on frequency.}
    \label{fig:narrow-band}
\end{figure}

We begin our calculation on the effectiveness of a broadband search by choosing the phase quadrature and  optimally tuning the target laser power, which corresponds to tuning the optomechanical coupling to a fixed target band centered at a frequency on the order of $1 \text{ MHz}$.  
We ensure that the coupling strengths for the momentum and the position sensing techniques are chosen such that for each technique the same target power is used during the broadband search. 
Figure~\ref{fig:broad-band} shows the potential benefits of squeezing while operating at the same power with respect to not using squeezing in the measurement. For the position sensing case, we are backaction limited at lower frequencies since we are operating at a high input power of $P\sim 1~\text{W}$. Such a high power is needed as compared to the $\nu = 10  \text{  kHz}$ case due to the higher frequency band-center. However, squeezing only provides a benefit at higher frequencies where the system is shot noise limited. The LIGO gravitational wave experiment encountered this limitation when they first introduced squeezing to their experiments ~\cite{aasi2013enhanced}.
For momentum sensing, at frequencies below the mechanical resonance $\omega_m$, backaction  dominates, so squeezing provides no improvement. The backaction noise decreases at frequencies above  the mechanical resonance $\omega_m$, where shot noise dominates. Hence, for momentum sensing, squeezing can significantly lower the noise in the regime $\nu > \omega_m$. Additionally, we see that momentum sensing is more advantageous than position sensing for broadband detection as it has a lower noise floor while operating at a fixed input power, especially for $\nu > \omega_m$.

Next, we consider the narrow-band search and calculate its sensitivity, as shown in Fig.~\ref{fig:narrow-band}. Here, we assume that the mechanical oscillator frequency is fixed. In the cases presented here it has been fixed at $\omega_m = 100 ~{\rm Hz}$.  We begin the search by tuning the input power $P$ at every frequency such that we set the noise  floor to the SQL for the scenario without squeezing across the frequency spectrum, as shown in Fig.~\ref{fig:narrow-band}(b). Simultaneously, at every frequency we select the optimal quadrature angle to eliminate the backaction noise. We compare the implementation of our narrow-band search with ($r=2$) and without ($r=0$) squeezing with equal power in both cases. Note that for the same power, squeezing lets us achieve a lower noise floor in both position and the momentum sensing. Here, by selecting the optimum quadrature the backaction term is eliminated and the remaining shot noise is reduced using squeezed light. For the position sensing case, we see that at frequencies below the mechanical frequency ($\nu \ll \omega_m$), the noise is proportional to $\hbar m \omega_m^2$ and independent of frequency. The noise remaining constant in this frequency range is due to selecting an optimal $G$. In the same regime for momentum sensing, the noise goes as $1/\nu^2$. The opposite is true for $\nu \gg \omega_m$ while operating at optimized powers, such that the noise is flat for  momentum sensing while it goes as $\nu^2$ for position sensing. This behavior  can be attributed to the fact that the optimized noise in the momentum sensing technique is related to the optimized noise in the position sensing technique by a factor of $\omega_m^2/ \nu^2$. Hence, for frequencies above the mechanical frequency we find momentum sensing to be more advantageous than position sensing. The roles reverse for frequencies below the mechanical frequency.  While we see no particular improvement in the minimum noise attainable on resonance, off resonance we do see a reduction in the noise floor by a constant factor dependent on the squeezing strength for both techniques. In another scenario, we could also benefit from squeezing in achieving the same noise floor at a lower power as discussed before.

\section{Outlook}

In this article, we show how to go beyond the SQL for  optomechanical sensors by leveraging squeezed light, backaction evasion, and QND techniques. These methods have previously been explored in the context of gravitational wave detection; however, their combination for enhanced force and impulse sensitivity has been somewhat under-appreciated in the existing literature. Thus, here we present a comprehensive overview of these techniques, especially in the context of the impulse metrology for dark matter detection. From our toy model and our cavity examples in the above sections \ref{section-single-sided-cavity} and \ref{BAE+squeezing}, we observe the benefit of squeezing in terms of the power requirements to achieve a given noise floor for position sensing. 
We also presented a  QND  momentum sensing protocol using a momentum coupling model, which shows similar benefits from squeezing in terms of power requirements when compared to the position sensing case. Critically, for the momentum sensing protocol in the free particle limit where the backaction term goes to zero, 
only the phase quadrature, which contains the information on the signal, needs to be measured over the full spectrum. Specifically, there is neither frequency nor power dependence to the optimal quadrature angle in the momentum sensing case, thereby reducing the technical challenges of using squeezing in comparison to squeezed position sensing. 
 
We have also explored search strategies for monochromatic and broadband signals. For broadband signals, we determined that our momentum sensing model provides a lower noise floor over a broad frequency range. For  both position and momentum sensing, the noise floor is lowered in the shot noise dominated regime using squeezed light at the same operating power as used in the case without squeezing. For monochromatic signals, in the low frequency regime, position sensing outperforms momentum sensing; while in the high frequency regime we see the opposite. Furthermore, by using squeezed light we can achieve a lower noise floor at all frequencies except on resonance for both protocols by optimally tuning the quadrature angle and power.

The quantum-enhanced sensing techniques we introduce are especially useful in the context of direct dark matter detection, as previously discussed \cite{carney2020proposal, carney2021ultralight}. Our techniques may also be used in impulse metrology experiments, such as sensing the collisions from background gas particles.  In order to  implement the momentum sensing technique, we can use specific designs of optomechanical systems or an electrical system for readout~\cite{richman2023general}.
Our analysis provides key insights that we may reduce the noise floor for an optomechanical sensor by combining backaction evasion, through a momentum measurement, and squeezing. Moreover, the momentum sensing approach has advantages in terms of bandwidth and robustness of the protocol at higher power. We foresee that deploying these techniques would aid in reducing the measurement-induced noise floor for optomechanical sensors and would pave the way for extremely sensitive impulse detection, beneficial to applications such as the search for dark matter.
 
\section{Acknowledgement}

We thank Peter Shawhan, Daniel Carney, and Jon Kunjummen for helpful conversations. SG is
supported by the Physics Frontier Center at the Joint
Quantum Institute, which is funded through the National
Science Foundation (Award no. 1430094).  Support for SG and JMT was provided by the U.S. DOE Office of Science, Office of High Energy Physics, QuantISED program (under FWP ERKAP63). This work was performed in part at Oak Ridge National Laboratory, operated by UT-Battelle for the U.S. Department of Energy under contract no. DE-AC05-00OR22725.  Support for RCP and postdoc support for SH was provided by the U.S. DOE Office of Science, Office of High Energy Physics, QuantISED program (under FWP ERKAP63).  Support for CEM, AMM, and postdoc support for MAF was provided by the U.S. DOE Office of Science, National Quantum Information Science Research Centers, Quantum Science Center.

\appendix
\section{Interaction Dependence on Displacement Amplitude in Single-Mode Squeezing Toy Model}
\label{appendix}
We consider continuously monitoring the position of a single-sided optomechanical system, comprised of a fixed mirror and a pendulous mirror. We probe the system with either classical or single-mode squeezed light as seen in the schematic diagram of Fig.~\ref{fig:singleModePos}(a) and the proposed experimental implementation of  Fig.~\ref{fig:experimental}(a). A typical single-sided optomechanical system, in the linearized regime where the operators for the fields in the cavity have been displaced by a large displacement amplitude $\alpha$ relative to the quadrature fluctuations~\cite{RevModPhys.82.1155, aspelmeyer2014cavity} has an optomechanical interaction described by the Hamiltonian 
\begin{equation}
    \begin{split}
        H_{\rm int} &= \hbar G  x X \text{,}\\
                    &= \hbar g \alpha  x X \text{.}
    \end{split}
\end{equation}
Here, $x$ is the position of the system, $X$ is the amplitude quadrature of the light, $g$ is the single-photon optomechanical coupling strength in frequency per unit length, and $G\equiv g \alpha $ is the optomechanical coupling strength enhanced by the displacement amplitude $\alpha$.  The displacement amplitude is proportional to the square root of the power of the probing light.  In this Hamiltonian, $\alpha$ sets the phase reference for our system and is assumed to be real ($\alpha \in \mathbb{R}$). By monitoring the phase of the light reflected from the system, we can determine the system position and  the backaction in the measurement. For a squeezed vacuum state, for example, this interaction Hamiltonian holds when the number of photons $n$ added to it by the displacement operation  is much greater than the squeezed vacuum photon number $n_{vac}$.

 For the unitary evolution of our toy models the interaction scales linearly by the factor $\zeta=\alpha\mu$. Here, the displacement amplitude $\alpha$ is amplified by the factor $\mu = g t_{int} \approx g / \kappa  $, where $g$ is the single-photon optomechanical coupling strength in frequency, $t_{int}$ is the interaction time, and $\kappa$ is the is the cavity linewidth. For an optomechanical cavity the intracavity photons are related to the displacement amplitude driving it by $\alpha=\sqrt n$. For typical single-photon optomechanical coupling strengths we consider $g  < 1$ kHz and cavity linewidths $\kappa > 10$ MHz~\cite{G_ref2_Nature_Wilson2015, G_ref3_Nature_Rossi2018}. As an example, near the minimum displacement amplitude we  drive the toy model with  ($\mu \alpha = 0.02$), the intracavity photon number is $n=4000$. The number of photons in a squeezed vacuum state is $\braket{n_{vac}} = \sinh^2{r}$. Hence, the photons generated by the displacement amplitude exceed the vacuum photon number $n_{vac}$ (with $r=2$ equivalent to squeezing of -17.37 dB)  by orders of magnitude, so we may neglect the contribution  of $n_{vac}$ to the displacement amplitude. 

\section{Two-Mode Squeezing Toy Model for Velocity Sensing}

Given our toy model in the main text describing position sensing with a two-mode squeezed set-up (as is depicted in main text in Fig.~\ref{fig:twoModePos}) we can now investigate the avenues to do velocity sensing with a two-mode squeezed set-up with some slight modifications to our previous design, aiming to benefit broadband measurements. For a velocity sensing set-up, we would like our readout system to pick up  information only about the momentum or velocity of the mechanical system but not its position. This is because the backaction from the probe gets incorporated into the position of the system and would affect subsequent measurements. For this reason, the most intuitive approach is to measure the sum of the phases of the two output modes, i.e. the phase sum quadrature, as the position information in the phase of each mode have opposite signs such that the signal would cancel. This can also be verified from Eq.~(\ref{two-mode-operators-evolution}). However, when taking the phase sum quadrature, the other information stored in the phase, including the desired mechanical momentum information, also cancels out.

Let us now consider a relative time delay between the interaction of the two modes of light with the mechanical system. In this model, the first mode will interact with the mechanical system at time $t_0$ and then the second mode will interact with it at time $t_1$. The free evolution of the system in between these two times is characterized by the unitary evolution $e^{-i \delta p^2/2}$ where $\delta$ is proportional to $t_1 -t_0$. The mechanical system then freely evolves for time $t_1-t_2$, characterized by the evolution operator, $e^{-i \beta p^2/2}$, until the first mode interacts again with the mechanical system at time $t_2$. The second mode interacts again with the mechanical system at time $t_3$ after a free evolution time of $t_3-t_2= t_1-t_0$. After the full evolution, 
\begin{equation}
    \begin{split}
         U_{\rm total} &= e^{i  x_0 \zeta X_2} e^{-i \delta \frac{p_0^2}{2}} e^{-i  x_0 \zeta X_1 } e^{-i \beta \frac{p_0^2}{2}} \\
         & e^{i  x_0 \zeta X_2} e^{-i \delta \frac{p_0^2}{2}} e^{-i  x_0 \zeta X_1 } \text{,}
    \end{split}
    \label{two-mode-velocity-unitary}
\end{equation}
the phase quadratures of the two modes take the following form
\begin{equation}
    \begin{split}
          U_{\rm total}^\dagger Y_1  U_{\rm total} &= Y_1- 2 \alpha x - \alpha (\beta + \delta) p \\
           & + \alpha^2 \beta (X_1-X_2) + \alpha^2 \delta X_1 \text{,}\\
           U_{\rm total}^\dagger Y_2  U_{\rm total} &= Y_2 +2 \alpha x 
         + \alpha (\beta + 3 \delta) p \\
         & - \alpha^2 (\beta+ \delta) (X_1-X_2) - 3 \alpha^2 \delta X_1 \text{.}
    \end{split}
    \label{two-mode-velocity-unitary}
\end{equation}
Upon investigating these equations we find that the phase sum quadrature does not have any position information, given that it takes the form
\begin{equation}
     Y_1^\prime+ Y_2^\prime = (Y_1+Y_2) +2 \alpha \delta p
          - \alpha^2 \delta (X_1-X_2) - 2\alpha^2 \delta X_1 \text{.}
\end{equation}
Here we can do a combination of quadrature measurements for phase sum $Y_1+Y_2$ and amplitude difference  $X_1-X_2$ quadratures, so that the backaction contribution from the amplitude difference quadratures is nullified. However, we will still be left with some amount of backaction from the $X_1$ quadrature. This is because after the first mode interacts, some amount of backaction gets introduced to the system in the evolution time before the second mode interacts and this effect can not be compensated for. To some extent this is comparable to our backaction evading double ring cavity resonator set-up here~\cite{ghosh2020backaction}. This toy model indicates how using two-mode squeezed light can be used for the purpose of velocity or momentum sensing with minimum quantum noise, though it doesn't fully remove all backaction related terms.

\medskip

\appendix

\medskip

\bibliographystyle{unsrt}
\bibliography{mybib}

\end{document}